\documentclass[iop]{emulateapj}
\usepackage[modulo]{lineno} % the modulo causes only one in 5 to be printed
\usepackage[svgnames]{xcolor}  %

\newcommand{\uv}{\mbox{$u$-$v$}}

\newcommand{\kms}{\mbox{km s$^{-1}$}}
\newcommand{\muas}{\mbox{$\mu$as}}
\newcommand{\muasyr}{\mbox{$\mu$as~yr$^{-1}$}}

\newcommand{\muJb}{\mbox{$\mu$Jy~beam$^{-1}$}}
\newcommand{\Ra}[4]{\mbox{${#1}^{\rm h} \; {#2}^{\rm m} \; {#3}\fs{#4} $}}
\newcommand{\dec}[4]{\mbox{${#1}\arcdeg \; {#2}\arcmin \; {#3}\farcs{#4} $}}
\newcommand{\rhoCSM}{\mbox{$\rho_{\rm CSM}$}}
\newcommand{\rhoeject}{\mbox{$\rho_{\rm eject}$}}

\newcommand{\thfl}{\mbox{$\theta_{\rm90\%\;flux}$}}

\newcommand{\scomp}{{\rm comp}}

\defcitealias{SN86J-1}{Paper I} 
\defcitealias{SN86J-2}{Paper II}

\shortauthors{Bietenholz \& Bartel}

\begin{document}

%\linenumbers
      
\title{SN 1986J VLBI. III. The Central Component Becomes Dominant}

\author{Michael F. Bietenholz\altaffilmark{1,2}}
\author{Norbert Bartel\altaffilmark{2}}

\altaffiltext{1}{Hartebeesthoek Radio Observatory, PO Box 443, Krugersdorp,
1740, South Africa}
\altaffiltext{2}{Department of Physics and Astronomy, York University, Toronto,
M3J~1P3, Ontario, Canada}

%\slugcomment{Version 8.0,   \today}
\slugcomment{Accepted for publication in the Astrophysical Journal}

\begin{abstract}
We present a new 5-GHz global-VLBI image of supernova 1986J, observed
in 2014 at $t = 31.6$~yr after the explosion, and compare it to
previous images to show the evolution of the supernova.  Our new image
has a dynamic range of $\sim$100 and a background rms noise level of
5.9~\muJb.  There is no significant linear polarization, with the
image peak being $<3$\% polarized.  The latest image is dominated by
the compact central component, whose flux density is now comparable to
that of the extended supernova shell.  This central component is
marginally resolved with a FWHM width of $900_{-500}^{+100}$~\muas,
corresponding to a radius of $r_\scomp = 6.7_{-3.7}^{+0.7} \times
10^{16}$~cm for a distance of 10~Mpc.  Using VLBI observations between
2002 and 2014, we measured the proper motions of both the central
component and a hot-spot to the NE in the shell relative to the quasar
3C~66A\@.  The central component is stationary to within the
uncertainty of 12~\muasyr, corresponding to 570~\kms.  Our
observations argue in favor of the central component being located
near the physical center of SN~1986J\@.  The shell hot-spot had a mean
velocity of $2810\pm750$~\kms\ to the NE, which is
consistent with it taking part in the homologous expansion of the
shell seen earlier.  The shell emission is evolving in a
non-selfsimilar fashion, with the brightest emission shifting inwards
within the structure, and with only relatively faint emission being
seen near the outer edge and presumed forward shock. An animation is
available in the electronic edition.
\end{abstract}


\section{Introduction}
\label{sintro}

\objectname[]{SN 1986J} was one of the most radio luminous supernovae
ever observed.  Its unusually long-lasting and strong radio emission
and its relative nearness makes it one of the few supernovae for which
it is possible to produce detailed images with very-long-baseline
interferometry (VLBI).  It is also one of the few supernovae still
detectable more than $t = 30$ years after the explosion, thus we have
been able to follow its evolution for longer than for any other SN for
which there are resolved VLBI images except for SN~1979C
\citep{SN79C-shell, Marcaide+2009a}.  It is one of the few SNe which
we have been able to follow observationally as it evolves towards a
supernova remnant, and it thus helps to build a connection between
supernovae and their remnants \citep[see][]{MilisavljevicF2017}.  We
continue here our series of papers on VLBI observations of SN~1986J:
\citet{SN86J-1, SN86J-2}, which we will refer to as Papers I and II
respectively, as well as \citet{SN86J-Sci}.

SN~1986J was first discovered in the radio, some time after the
explosion \citep{vGorkom+1986, Rupen+1987}. The best estimate of the
explosion epoch is $1983.2 \pm 1.1$ \citep[Paper I, see
  also][]{Rupen+1987, Chevalier1987, WeilerPS1990}\nocite{SN86J-1},
which we take as $t = 0$.  It occurred in the nearby galaxy NGC~891,
for whose distance the NASA/IPAC Extragalactic Database (NED) lists 19
measurements with a mean of $10.0 \pm 1.4$~Mpc, which value we adopt
throughout this paper.

Optical spectra, taken soon after the discovery, showed a somewhat
unusual spectrum with narrow linewidths, but the prominent H$\alpha$
lines led to a classification as a Type~IIn supernova
\citep{Rupen+1987}.
Due to its relatively high radio flux density, it was one of the first
SNe to be observed with VLBI \citep{Bartel+1987}.  A VLBI image, the
first of any optically identified supernova, was obtained by
\citet{Bartel+1991}.  VLBI observations at subsequent epochs up to
2008 led to a series of images.  The source morphology was complex,
showing an expanding, albeit somewhat distorted shell with a prominent
hot-spot in the shell visible once the resolution was sufficient.

Other than the relatively slow turn-on, and its high radio luminosity,
the evolution of SN~1986J's spectral energy distribution (SED) was
unremarkable till 1998\@.  In \citetalias{SN86J-1}, we showed that at
that time, an inversion appeared in the spectrum, with the brightness
increasing with increasing frequency above $\sim$10~GHz, up to a
high-frequency turnover at $\sim$20~GHz.

In \citet{SN86J-Sci}, we showed by means of phase-referenced
multi-frequency VLBI imaging that this spectral inversion was
associated with a bright, compact component in the projected center of
the expanding shell. Such a central component has so far not been seen
in any other supernova\footnote{We note that central emission at mm
  wavelengths has been seen in SN~1987A, which is attributed to dust.
  No central synchrotron emission has been seen in SN~1987A at cm
  wavelengths \citep{Zanardo+2014}.}
\citep[see e.g.,][]{SNVLBI_Cagliari, BartelB2014IAUS}.  At that time,
in late 2002, the central component was clearly present in the 15~GHz
image, but not discernible in the 5~GHz one.  

In \citetalias{SN86J-2} we showed that from 2005 on, the central
component had become bright also at 5~GHz.  We also showed that,
though the morphology was complex, the radio emission region was still
expanding but also decelerating, with the average outer radius $\propto
t^{0.69 \pm 0.03}$.

\section{Observations and Data Reduction}
\label{sobs}

\subsection{VLBI Observations}
\label{svlbi}

We obtained VLBI observations of SN~1986J on 2014 Oct.\ 22, using a
global VLBI array which consisted of the following 21 antennas.  From
the National Radio Astronomy Observatory (NRAO)\footnote{The National
  Radio Astronomy Observatory, NRAO, is a facility of the National
  Science Foundation operated under cooperative agreement by
  Associated Universities, Inc.}:
the VLBA ($9 \times 25$~m diameter, Mauna Kea did not observe), the
Jansky Very Large Array (in phased mode, equivalent diameter 130~m);
from the European VLBI Network: Badary (32~m), Effelsberg (100~m
diameter), Hartebeesthoek (26~m), Jodrell Bank Lovell (70~m), Noto
(32~m), Onsala (25~m), Svetloe (32~m), Torun (32~m), Westerbork
(phased mode, equivalent diameter 94m), Yebes (40~m) and
Zelenchukskaya (32~m).

The observations of SN~1986J were interleaved with ones of the quasar
\objectname[]{3C~66A}, only 40\arcmin\ away on the sky, which we used
as a phase-reference source.  SN~1986J's declination of
+42\arcdeg\ enabled us to obtain dense and only moderately elliptical
\uv~coverage.  As usual, a hydrogen maser was used as a time and
frequency standard at each telescope. We recorded both senses of
circular polarization with the RDBE/Mark5C wide-band system at a
sample-rate of 1~Gbps, and correlated the data with NRAO's VLBA DiFX
correlator \citep{Deller+2011a}.  We used a bandwidth of 128~MHz
centered on 4.996~GHz.
The data reduction was carried out with NRAO's Astronomical Image
Processing System (AIPS).  The initial flux density calibration was
done through measurements of the system temperature at each telescope,
and improved through self-calibration of the 3C~66A data.

We used a cycle time of $\sim$3.2 minutes to phase-reference to
3C~66A, with about 1.8 minutes spent on SN~1986J\@.  Our positions in
this paper are given relative to an assumed position of the brightness
peak of 3C~66A of RA = \Ra{02}{22}{39}{611500}, decl.\ =
\dec{43}{02}{07}{79884} taken from the International Celestial
Reference Frame, ICRF1\footnote{We note that a slightly revised
  position for 3C~66A is given in the ICRF2 \citep{FeyGJ2009}, which
  is different from the one we used by $-43, -24$~\muas\ in RA and
  decl.\ respectively \citep{Ma+1998}.  However, for consistency, we
  use in this paper the same ICRF1 reference position for 3C~66A as we
  used in \citetalias{SN86J-2}, which has no effect on the differential
  positions or any of our conclusions.}.

We determined the instrumental polarization leakage from our
observations of 3C~66A, which is almost unpolarized, using a linear
approximation (AIPS task LPCAL). We did not calibrate absolute
position angle, so our polarization measurements resulted in correct
magnitudes but unknown position angles.

\subsection{VLBI Imaging Considerations}
\label{simaging}

Imaging and deconvolving interferometer data can be difficult,
particularly in the case of a non-uniform and relatively sparse array
like that used for VLBI.  We use the multi-scale extension of the
original CLEAN algorithm, MS-CLEAN \citep{WakkerS1988}, for the
deconvolution.  MS-CLEAN works similarly to the traditional CLEAN
algorithm \citep[see, e.g.,][]{CornwellBB1999}, but instead of just
using a single basis function based on the central lobe of the dirty
beam, that is one based on the ``native'' or highest resolution
possible in the data, MS-CLEAN simultaneously deconvolves images at
that native resolution and several larger ones, and then combines the
different resolutions into a final image.  MS-CLEAN has generally been
shown to produce superior results for the deconvolution of extended
sources \citep[see, e.g.,][]{Hunter+2012, Rich+2008, SNR4449}.

\citet{GreisenSvM2009} and \citet{Cornwell2008} discuss the properties
of MS-CLEAN as implemented in AIPS.  Aside from the selection of
resolutions, the main parameters for tuning the MS-CLEAN algorithm are
the order in which the different resolutions are treated, and the
depth to which they are CLEANed \citep{GreisenSvM2009}.  We made two
different simulated visibility data sets for testing, both designed to
be relatively similar to the brightness distribution of SN~1986J.  The
first one was a geometrical model consisting of an almost circular
disk of diameter $\sim$8~mas and a slightly off-center
marginally-resolved Gaussian source with full-width at half-maximum
(FWHM) 0.4~mas.  The second model consisted of the CLEAN components
derived from a deconvolution of the real data.  We replaced the
observed visibility values with the relevant Fourier transforms of
these models and then added noise at a level comparable to that in the
observed visibilities.  We deconvolved the resulting simulated
visibilities, and compared the resulting CLEAN images to the model
images after convolving the latter with the CLEAN beam.

We found that the most accurate reconstruction was achieved for both
models when the CLEAN iterations were stopped when the flux density of
the CLEAN components had reached the rms image background level. We
also found that a multi-scale CLEAN with three resolutions, with FWHM
beam areas of 2.2, 10.5 and 36.1 mas$^2$ gave good results.
Using this imaging strategy, the rms error within the CLEAN window,
i.e., the difference between the CLEANed image and the original
model, was only 3\% and 16\% larger than the rms image background for
the two models respectively, with the worst-case errors being
$\lesssim 5\times$ the image background rms.

\section{VLBI Images}  
\label{svlbiimg}

\begin{figure*}
\centering
\includegraphics[width=0.7\linewidth]{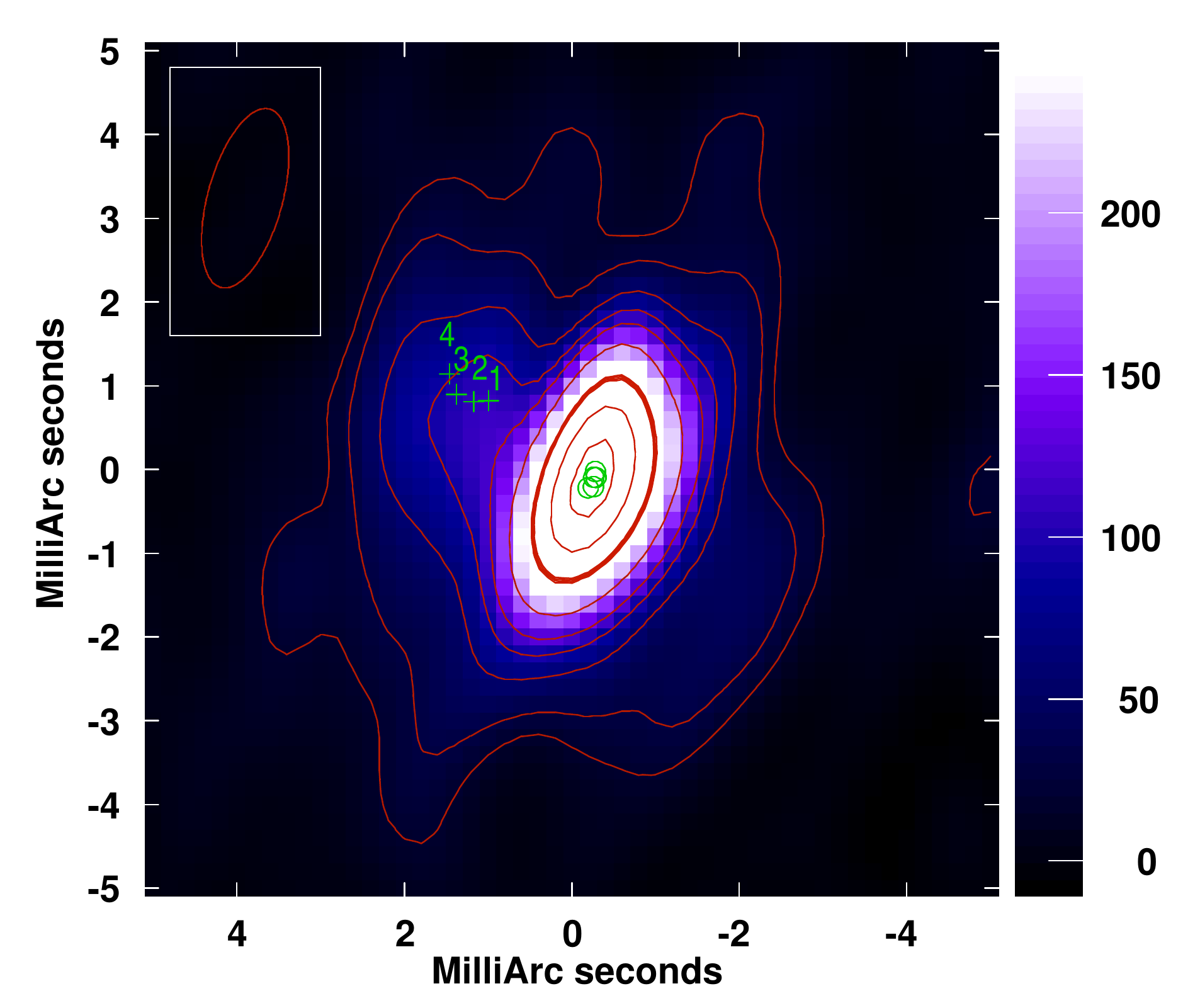}
%C  SN86JOC14PS3.ICL001. 36; (self-cal from SN86JOC14FXW.C12FXW.2)
\caption{The 5-GHz VLBI image of SN~1986J made from observations on
  2014 Oct.\ 23, at age 31.6~yr.  The contours are drawn at $-3$, 3,
  5, 10, 15, 20, 30, {\bf 50}, 70 and 90\% of the peak brightness,
  with the 50\% contour being emphasized.  The peak brightness was
  617~\muJb, the total CLEANed flux density was 1622~$\mu$Jy, and the
  background rms brightness was 5.9~\muJb.  The color or gray scale is
  labeled in \muJb, and is saturated so as to better show the
  low-level emission.  The (green) overlapping circles indicate the
  position of the central component at ages 20.3, 22.6, 25.6 and
  31.6~yr, while the (green) crosses labeled ``1'' to ``4'' show the
  position of the shell hot-spot at ages 15.9, 19.6, 22.6, and
  25.6~yr, respectively, along with their estimated uncertainties of
  120~\muas\ in each coordinate. In the present
  image ($t = 31.6$~yr), the shell hot-spot is no longer clearly
  identifiable.  North is up and east to the left, and the FWHM of the
  convolving beam of 2.21~mas $\times 0.89$~mas at
  p.a.\ $-15$\arcdeg\ is indicated at upper left.}
\vspace{0.1in} % kluge to improve spacing
\label{fimg}
\end{figure*}

We show the 5-GHz VLBI image of SN~1986J in Figure~\ref{fimg}.  The
image was made in AIPS, using complex weighting with the robustness
parameter chosen so as to attain close to natural weighting (AIPS
ROBUST = 2.2).  Furthermore, we used multi-scale CLEAN deconvolution
with the strategy discussed in \S~\ref{simaging} above.  The total
CLEANed flux density was 1622~$\mu$Jy, the peak brightness 617~\muJb,
and the background rms brightness was 5.9~\muJb, for an image dynamic
range of $\sim$100.  We estimate that the on-source brightness
uncertainty is $\sim$7~\muJb, only slightly larger than the off-source
rms.

The image is dominated by the central component.  In contrast, the
highest brightness in the more extended emission is only $\sim
100$~\muJb, or $\lesssim$15\% of the peak brightness of the central
component (we estimate the brightness ratio between the central
component and the remainder below in \S~\ref{sfluxd}).

\begin{figure*}
\centering
\includegraphics[height=0.90\textheight]{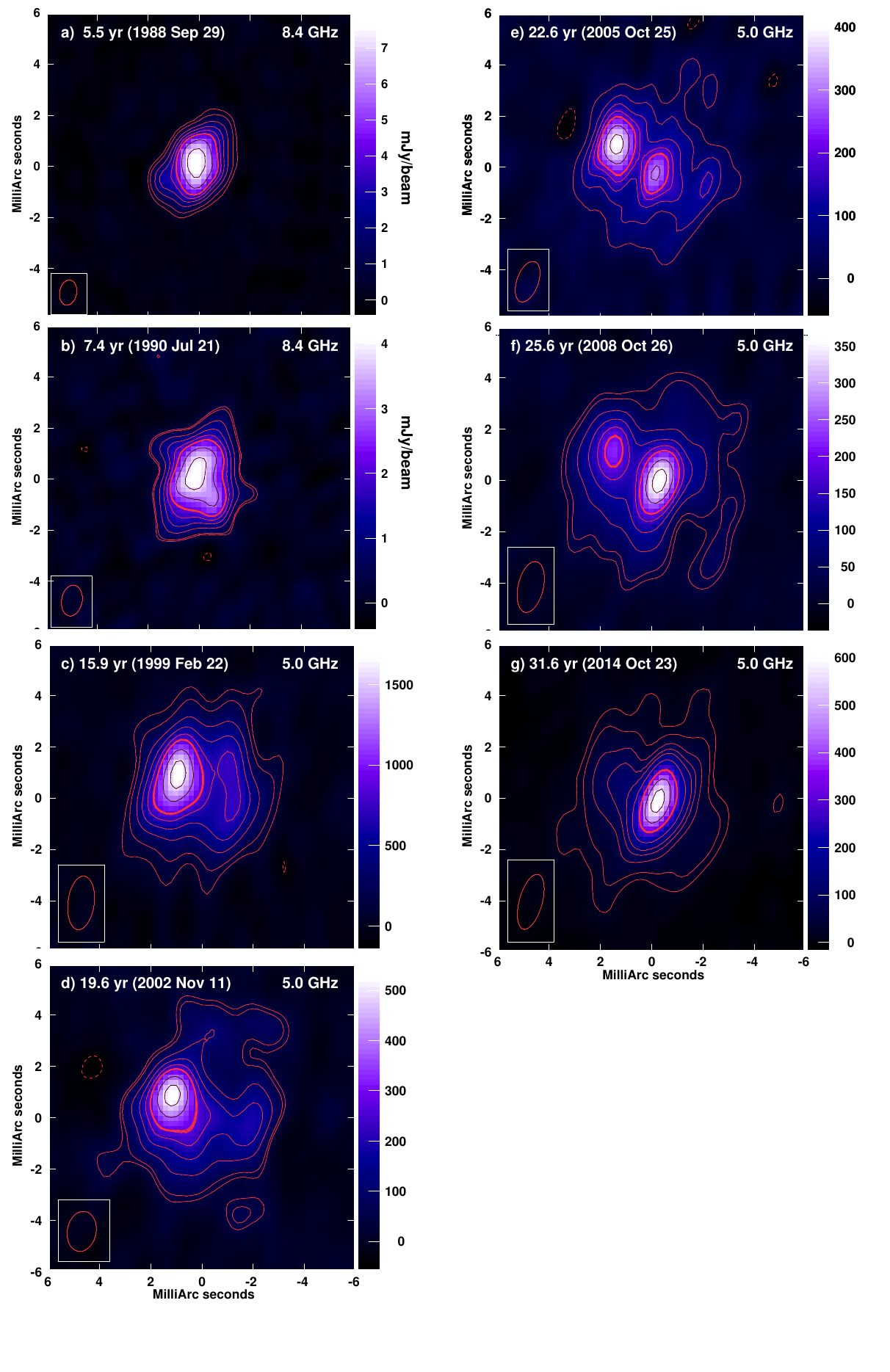}
\caption{\small A sequence of all our VLBI images of SN~1986J at 5 or
  8.4~GHz, showing its evolution from $t = 5.5$~yr to 31.6~yr after
  the explosion, or 1988 to 2014\@.  Each panel has the lowest contour
  at $3\times$ the image rms, and the remainder at 10, 20, 30, 40,
  {\bf 50}, 70, and 90\%, of the peak brightness, with the 50\% one
  being emphasized (and being the first white contour in the black and
  white images). In each panel, the age and date are given at top
  left, the frequency at the top right, and the FWHM size of the
  convolving beam is indicated at lower left.  North is up and east to
  the left. The gray- or colorscale is in \muJb\ unless indicated
  otherwise. For details of panels {\em a)} and {\em b)}, see
  \citetalias{SN86J-1}, {\em c)} through {\em f)}, see
  \citetalias{SN86J-2}.  Panel {\em g)} is the same as
  Figure~\ref{fimg}, but not saturated.  This figure is also available
  as an animation in the electronic edition.}
\label{fimgseq}
\end{figure*}

To highlight the evolution of SN~1986J, we show in
Figure~\ref{fimgseq} our earlier images at 8.4 GHZ, reproduced from
\citetalias{SN86J-1}, and at 5.0 GHz, reproduced from
\citetalias{SN86J-2}.
In the first image, at $t = 5.5$~yr and 8.4~GHz, the source is still
rather compact but with the protrusions already visible. By $t =
7.4$~yr the protrusions had expanded indicating a complex brightness
distribution.
At $t=15.9$~yr and at 5 GHz, the shell hot-spot is visible to the NE
of the center.  By $t = 22.6$~yr, the central component is starting to
appear, and by $t = 25.6$~yr, the central component has become
brighter than the shell hot-spot, which is fading.  By the present
epoch, at $t = 31.6$~yr, the central component dominates and the shell
hot-spot is no longer clearly visible.

\subsection{Polarization}
\label{spoln}

We also made images in Stokes Q, U, and V.  As mentioned in section
\ref{svlbi} above, we did not calibrate the absolute polarization
p.a., but we did calibrate the leakage terms.  Therefore, our
measurements of the linearly polarized intensity are accurate, but
that of the p.a.\ are completely uncertain.  The lack of
p.a.\ calibration turns out to be immaterial, since no significant
polarized emission is seen in either linear or circular polarization.
We combined the Q and U images into an image of the linearly polarized
intensity, $S_{\rm pol}$, corrected for the Ricean noise
bias\footnote{$S_{\rm pol} = \sqrt{Q^2 + U^2}$ where $Q$ and $U$ are
  the measured Stokes parameters.  By definition, $S_{\rm pol}$ is
  positive, and in the presence of noise is therefore biased.  This
  bias has been calculated and a bias correction has been implemented
  in the AIPS task COMB, which we used to calculate our linearly
  polarized flux densities.  This procedure produces the correct mean
  value of $S_{\rm pol}$, at the expense of occasionally producing
  unphysical negative values.}.
The rms in Stokes Q and U was 4.4~\muJb, while that in $S_{\rm pol}$
was 4.6~\muJb.  No significant polarization was detected, with the
maximum value of $S_{\rm pol}$ observed being 14~\muJb\ or $3\sigma$.
In particular the polarized intensity at the location of the central
component was $7.0 \pm 5.6$~\muJb, and we put a $3\sigma$ upper limit
on the linear polarization of the central component of 3.3\%.  This
limit is comparable to upper limits we reported for SN~1993J, and
consistent with the expectation that internal Faraday depolarization
is quite strong \citep{SN93J-3}.

\subsection{The Central Component}
\label{scentral}

We determined the characteristics of the central component by fitting
an elliptical Gaussian model as well as a zero level to the center of
the CLEANed (Stokes I) image by least squares.  Although the properties
of the fitted Gaussian are somewhat dependent on the exact choice of
the fitting window because the central component cannot be separated
uniquely from the complex background emission due to the shell, the
results can nonetheless be used to characterize the central component.

\pagebreak[4]
\subsubsection{Flux Density}
\label{sfluxd}

The best fitting point source model, i.e., an elliptical Gaussian of
width fixed to that of the CLEAN beam, had a flux density of $511 \pm
5 \; \mu$Jy.  
We note that the value of $511\; \mu$Jy is, strictly speaking, closer
to being an estimate of the peak brightness in \muJb\ than an estimate
of the total flux density, since if the component is resolved the
latter would be higher.  We do in fact find that the central component
is slightly resolved (see \S~\ref{swidth} below).  However, it is
difficult to reliably estimate both the width and the flux density
simultaneously.  In order to be able to compare the flux density from
this epoch with those from previous ones consistently, we therefore
take the values obtained by fitting an unresolved source and a zero
level to the image as the estimates of the flux density, keeping in
mind that we are underestimating the total flux density to the extent
that the central component is resolved.

This value of the flux density of the central component at $\nu =
5$~GHz corresponds to a spectral luminosity, $L_\nu$, of $6 \times
10^{25}$ erg~s$^{-1}$~Hz$^{-1}$, or $\sim$20 times that of the Crab
Nebula.  If we approximate the luminosity as $\nu L_\nu$, the central
component's luminosity is $\sim 80 \; L_{\sun}$.

\begin{figure}
\centering
\includegraphics[width=\linewidth]{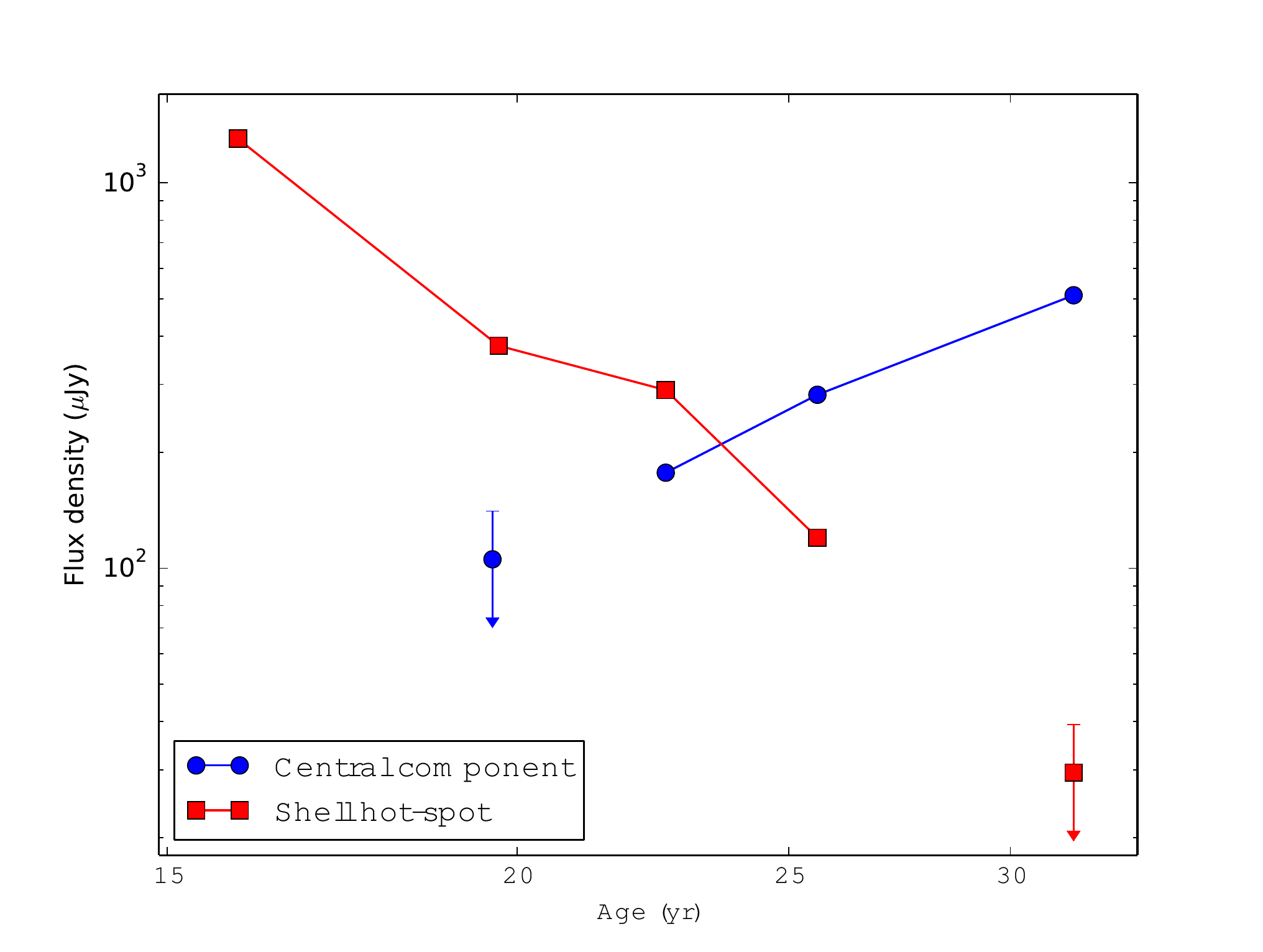}
\caption{The 5-GHz flux densities of the central component (blue
  circles) and the shell hot-spot (red squares) as determined from the
  VLBI images (see text), as a function of age.  For the epochs at
  which the component or spot could not be clearly discerned, we plot
  50\% of the surface brightness at the relevant location as the upper
  limit, on the grounds that if the feature had more than half the
  brightness of the background emission at that location, it would be
  clearly discernible in the image.
  % kluge to add a bit of space here.
  \vspace{0.1in}}
\label{fspotflux}
\end{figure}

If we fit the images from earlier epochs in the same way, we obtain
the values tabulated in Table~\ref{tspot}.  The uncertainty in these
flux-density estimates is difficult to determine accurately as it
depends on the separation between the central component and the
shell-emission in the fit and how resolved the central component is.
However, the general picture seems clear: from $t = 22.6$~yr to $t =
31.6$~yr (2005 to 2014) the central component's flux density {\em
  increased}\/ by $\sim 300 \; \mu$Jy (from 177 to 511 $\mu$Jy), or by
a factor of almost 3.
As a fraction of SN~1986J's total flux density, it has increased from
about 7\% at $t = 22.6$~yr to 32\% at $t=31.6$~yr.  We can therefore
say with reasonable confidence that, in the last 6 yr, the central
component's fraction of SN~1986J's total 5-GHz flux density has
roughly doubled, and that the 5-GHz flux density of the central
component has increased.  We plot the flux density of the central
component in Figure~\ref{fspotflux}.

\begin{deluxetable*}{c c r c c c c c c}
\tablewidth{0pt}
\tablecaption{Flux densities and positions offsets of the central
  component and the shell hot-spot \label{tspot}}
\tablehead{
\colhead{Midpoint Date} & \colhead{Age\tablenotemark{a}} & 
\colhead{Frequency} & \multicolumn{3}{c}{Central component} & 
\multicolumn{3}{c}{Shell hot-spot} \\
 & & & \colhead{Flux density\tablenotemark{b}} & 
\multicolumn{2}{c}{Position offset\tablenotemark{c}}  & 
\colhead{Flux density\tablenotemark{b}} & 
\multicolumn{2}{c}{Position offset\tablenotemark{c}} \\
 & & & & \colhead{RA} & \colhead{dec.} & & \colhead{RA} & \colhead{dec.} \\
  & \colhead{(yr)}  & \colhead{(GHz)} &
    \colhead{($\mu$Jy)} & \colhead{(mas)} & \colhead{(mas)} &
    \colhead{($\mu$Jy)} & \colhead{(mas)} & \colhead{(mas)} }
\startdata
 1999 Feb 22 & 15.94 &  5.0 &\nodata& \nodata   &  \nodata  &1300\phn& 0.996 & 0.821 \\ %start 21 Feb
 2002 Nov 11 & 19.66 &  5.0 &\nodata& \nodata   &  \nodata  &  380   & 1.172 & 0.809 \\ %start 10 Nov
 2003 Jun 22 & 20.27 & 15.4 &\nodata& $-0.261$ & $-0.207$ & \nodata&\nodata&\nodata\\ %start 21 Jun
 2003 Jun 23 & 20.28 &  8.4 &\nodata& $-0.289$ & $-0.097$ & \nodata&\nodata&\nodata\\ %start 22 Jun
 2005 Oct 25 & 22.62 &  5.0 & 177  & $-0.195$ & $-0.221$  & 290 & 1.376 & 0.895 \\ %start 24 Oct
 2008 Oct 26 & 25.62 &  5.0 & $\;282$\tablenotemark{d}  
                              & $-0.283$ & $-0.025$  & 120 & 1.460 & 1.143 \\ %start 25 Oct
 2014 Oct 23 & 31.61 &  5.0 & 511  & $-0.263$ & $-0.095$  & \nodata&\nodata&\nodata\\
\enddata
\tablenotetext{a}{The age of SN 1986J, taken with respect to an explosion
epoch of 1983.2 \citepalias[see][]{SN86J-1}.}
\tablenotetext{b}{The flux density of an unresolved source fitted to
  the image.  A variable zero-level was simultaneously fitted.  This
  fit will underestimate the total flux density of a resolved source,
  but we use this consistent estimate to compare the various epochs,
  since we cannot reliably determine both the extent and the flux
  density.}
\tablenotetext{c}{The position offset from the our
    nominal explosion center position.  We use the interpolated
    peak-brightness location from the un-selfcalibrated 5-GHz images
    as our estimate of the positions, and the estimated uncertainties
    are 120~\muas\ in each coordinate.  Our nominal explosion center
    position is the average position of the center of the shell from
    1999 to 2008, RA = \Ra{02}{22}{31}{321457}, decl.\ =
    \dec{42}{19}{57}{25951}, and we estimate that this position is
    within $\sim$200~\muas\ of the true explosion center
    \citepalias[see][]{SN86J-2}.}
\tablenotetext{d}{Note that in \citetalias{SN86J-2} we give an
  approximate and somewhat larger flux density of 390~$\mu$Jy for the
  central component which was derived just from the peak brightness in
  the image.  Fitting the zero-level, as we do here, gives a somewhat
  lower, but more accurate estimate of the flux density of the
  component.}
\end{deluxetable*}

The peak brightness of the central component, as estimated by the
point-source fit, is $511 \pm 5$~\muJb.  The average brightness of the
shell emission is $\sim$50~\muJb, comparable to, but slightly lower
than, the shell-brightness of $70\pm8$~\muJb\ estimated by our fit at
the central component location.  The central component is therefore
$\sim 10 \times$ brighter than the shell is on average.  

The shell emission is optically thin, since it has a spectral index of
$\sim -0.6$ (where $S_\nu \propto \nu^\alpha$).  The central
component, on the other hand, had a positive spectral index at 5~GHz
at $t \sim 25$~yr and was therefore still optically thick, with the
transition to an optically thin spectrum occurring near 13~GHz
\citepalias{SN86J-2}.  Extrapolating the SED from \citetalias{SN86J-2}
to $t = 31.6$~yr, it seems unlikely that the SED would have evolved to
the point it was optically thin at 5~GHz, so the central component is
almost certainly still partly absorbed at 5~GHz.  The intrinsic
brightness ratio between the central component and the shell is
therefore likely even higher than 10.

\pagebreak[4]
\subsubsection{Position and Proper Motion}
\label{sposition}

We take the quadratically interpolated peak brightness position from
the phase-referenced image as the best estimate of the central
component's position at $t = 31.6$~yr, which was RA =
\Ra{02}{22}{31}{321434} and decl.\ = \dec{42}{19}{57}{25941}
with small statistical uncertainties of $< 10$~\muas.  A fit of an
elliptical Gaussian to the image results in a position consistent with
the one given above to within 20~\muas.

For a realistic estimate of the uncertainty on the position which
includes systematic effects, we take the same conservative estimate of
120~\muas\ as we did in \citetalias{SN86J-2}.  We confirmed this
estimate with a Monte-Carlo simulation based on the present data.  The
largest part of this uncertainty is due to the difficulty in
separating the central component from the shell emission as well as
due to noise, with a smaller part being due to the possible
instability in the reference source and errors in the
phase-referencing such as the troposphere or errors in the station
positions \citep[for estimates of the phase-referencing errors,
  see][]{PradelCL2006}.  We tabulate the position of the central
component, expressed as offsets from our nominal
explosion center position, in Table \ref{tspot}.

For our nominal explosion center position, we took the estimate of the
explosion center location that we found in \citetalias{SN86J-2}, which
was RA = \Ra{02}{22}{31}{321457}, decl.\ = \dec{42}{19}{57}{25951},
with an uncertainty of 200~\muas\ in each coordinate.  This estimate
is the position of the center of the shell, averaged over 5 epochs
from 1999 to 2008, all determined with respect to 3C~66A.  If the
supernova is expanding symmetrically, this shell center position is
identical to the explosion center.  Only in two supernovae do we have
observational constraints on the symmetry of the ejecta in the sky
plane.  In SN~1993J, in projection, the ejecta are circularly
symmetric to within 5.5\% about the explosion position
\citep{SN93J-1}.  In SN~1987A, there is a complicated structure, with
mostly bilateral symmetry, but with one-sided asymmetry of at least
10\% at some azimuth angles \citep{Zanardo+2013}.
For SN~1986J, the presence of two distinct blue-shifted components
in the optical spectra also suggest an asymmetric structure
\citep{Milisavljevic+2008}.

In 2014, therefore, SN~1986J's central component is $-263,
-95$~\muas\ from the explosion center in RA and dec.\ respectively,
with a combined uncertainty of 230~\muas\ in each coordinate.  We
regard this displacement as suggestive, but not conclusive.

\begin{figure}
\centering
\vspace{0.2in}  % need this kluge to avoid overlapping the table....
\includegraphics[width=\linewidth]{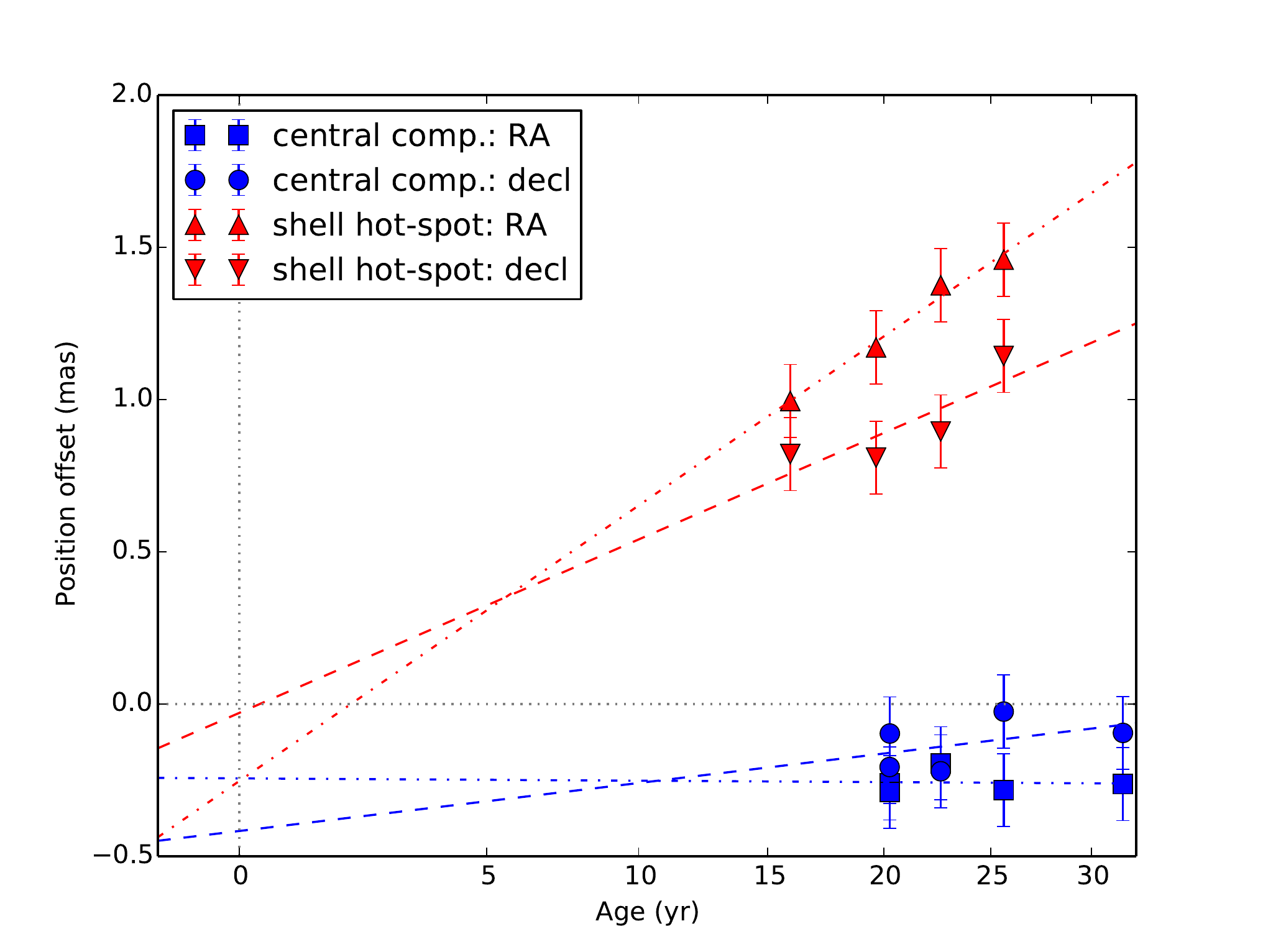}
\caption{The proper motion of the central component and the shell
  hot-spot.  The ``Age'' axis, $t$, is plotted non-linearly, so that
  any offset $\propto t^{0.69}$ would appear as a straight line in the
  plot.  We choose this scaling because we determined in
  \citetalias{SN86J-2} that the shell expands with radius $\propto
  t^{0.69}$.  The positions were determined from images
  phase-referenced to 3C~66A\@, and are listed in Table~\ref{tspot},
  along with the frequency of each image.  The position offsets are
  determined relative to 3C~66A, but are expressed relative to RA =
  \Ra{02}{22}{31}{321457}, decl.\ = \dec{42}{19}{57}{25951}), which is
  the average position of the center of the shell
  \citepalias[see][]{SN86J-2}. The plotted $1\sigma$ uncertainties of
  120~\muas\ are estimates which include both statistical and
  systematic contributions added in quadrature.  The plotted lines
  indicate the fits of the form $x = x0 + b*t^{0.69}$, with the
  dot-dashed lines being RA and the dashed lines being
  decl. \vspace{0.1in} }
\label{fspotposn}
\end{figure}

We can determine the proper motion of the central component by
comparing the position in the present epoch with
consistently-determined ones from 2005 and 2008 \citepalias{SN86J-2},
which are tabulated in Table~\ref{tspot}.  A linear regression gives a
proper motion of $0 \pm 12$ and $8 \pm 12$ \muasyr\ in RA and decl.,
respectively, corresponding to $380 \pm 570$~\kms\ at p.a.\ $\sim
0$\arcdeg.  We plot the proper motion of the central component in
Figure \ref{fspotposn}.

\subsubsection{Size and Expansion Velocity}
\label{swidth}

We also allowed the fitted model to be non point-like by allowing the
widths and p.a.\ of the fitted elliptical Gaussian to be free in the
fit.  We obtained a fitted FWHM, deconvolved from the CLEAN beam, of
$900 \times 550$~\muas\ at p.a.\ 125\arcdeg, a flux density of
707~$\mu$Jy, and a fitted zero-level of 45~\muJb.
The rms residuals to this fit were 16~\muJb, about 2.7 times larger
than the off-source background rms. Unfortunately, the fitted
deconvolved size is biased when it is near 0, since it cannot be less
than 0\@.  We performed a Monte-Carlo simulation ($n=400$~trials),
where we used model sources of known size and rms image-noise levels
of 16~\muJb\ (equal to the rms residuals in our fit to the real
image). This simulation showed us that there is a 4\% chance that we
would obtain a fitted major axis of 900~\muas\ when the source was in
fact unresolved, resulting in final values for the fitted FWHM and
p.a., along with the corresponding uncertainties, of
$900_{-500}^{+100}$~\muas, at p.a.\ $125\arcdeg \pm25\arcdeg$\@.
In the NS direction, our resolution is poorer, and we can only say
that the component's perpendicular axis is $\lesssim 1000$~\muas,
therefore an approximately circular shape is compatible with our
measurements.  These FWHM values can be compared to our FWHM
resolution, which was $1710 \times 760$~\muas\ at p.a.\ $-15$\arcdeg.
The residuals to the fit were, as mentioned, $\sim 2.7\times$ larger
than the expected image noise.  This is likely due to the shell
emission in our fit being represented only by a constant offset, which
is almost certainly an over-simplification.  It could also be due to
the central component being resolved and having a shape different from
an elliptical Gaussian.

Our fitted value of $900_{-500}^{+100}$~\muas\ for the FWHM major axis
of the central component suggests that the central component is
somewhat resolved in an approximately SE-NW direction.  Both the fitted
FWHM width and p.a.\ are consistent with what can be seen in the
higher resolution 22-GHz image in \citetalias{SN86J-2}.

At 10~Mpc, the major axis FWHM corresponds to $r_{\rm comp} =
6.7_{-3.7}^{+0.7} \times 10^{16}$~cm.  The average expansion velocity,
over the 31.6~yr, assuming the central component originated in the SN
explosion, is then $680^{+80}_{-380}$~\kms.

\subsection{The Shell Hot-Spot}
\label{sshellspot}

We first noted a bright hot-spot to the NE in the shell in
\citetalias{SN86J-1}, where we called it C1\@.  In our current image,
at $t=31.6$~yr (Fig.~\ref{fimg}), the shell hot-spot is no longer
clearly discernible, although there is still an enhancement of the
brightness near the location where the shell hot-spot is expected.
The shell hot-spot has been fading steadily since we first noted it.
We give estimates of its 5-GHz flux density, or limits thereon in the
last epoch, determined in the same way (\S~\ref{sfluxd}) as were those
of the central component (and which will similarly underestimate the
true flux density if the shell hot-spot is in fact resolved).  We plot
the flux densities and limits in Figure~\ref{fspotflux}.

We tabulate the shell hot-spot's position, again expressed as offsets
from our nominal explosion center position, in Table \ref{tspot}. The
positions, like those of the central component (see
\S~\ref{sposition}), were the quadratically interpolated peak
brightness positions on the phase-referenced images).  We plot the
positions(along with those of the central component) in
Figure~\ref{fspotposn}.  Since we cannot determine a position for the
hot-spot in the present image, we have only the same set of positions
that we already determined a proper motion from in
\citetalias{SN86J-2}.  However, a new analysis (using a weighted fit)
gives a consistent but slightly more accurate proper motion of $59 \pm
16$~\muasyr, corresponding to a projected velocity of $2810 \pm
750$~\kms, at p.a.\ $(57 \pm 15)$\arcdeg.

As we noted in \citetalias{SN86J-2}, a decelerated motion like that
seen for the shell, with $r \propto t^{0.69}$, is also consistent with
the measured positions and we show a fit of this form in
Figure~\ref{fspotposn}.  The shell hot-spot therefore, has a radially
outward proper motion consistent with the homologous expansion of the
shell, with a projected speed about half that of the outer edge of the
radio emission, consistent with its projected position about halfway
between the average outer radius and the center.  Such a proper motion
would be expected if it were due to a dense condensation in the CSM.

We note that the p.a.\ of the shell hot-spot is approximately in the
same direction as the elongation of the central component which is
visible in our 2006 image at 22~GHz \citepalias{SN86J-2}.  Furthermore
there is a suggestion in some of the images in Fig.~\ref{fimgseq} of a
possible enhancement on the SW side, opposite the shell hot-spot. We
elaborate on this coincidence in the Discussion section
(\S~\ref{sdiscuss}).

\pagebreak[4]
\subsection{The Shell Size and Expansion Curve}
\label{sexpand}

In our earlier papers, we determined the outer radius of SN~1986J at
each epoch, and determined the corresponding expansion curve.  We
found that the outer radius was expanding with time as $r_{\rm out}
\propto t^{0.69 \pm 0.03}$ \citepalias{SN86J-2}.  Can we again
estimate the angular radius reliably for our new observations to track
the expansion curve?

In our earlier papers, we took as a representative value for the
angular outer radius of SN~1986J the value \thfl, which is equal to
$\sqrt {\mathrm {area}/\pi}$ of the contour which encompasses 90\% of
the total flux density.  Since \thfl\ is somewhat dependent on the
convolving beam size, we convolved our images with an approximately
co-moving beam whose size increases $\propto t^{0.70}$ so as to
minimize the bias in the measured radius evolution.

We convolved our new, $t = 31.6$~yr, image with the co-moving beam
(FWHM: 3.91 mas $\times 1.96$ mas at p.a.\ $-1$\arcdeg), and found
that the formal value of \thfl\ is now 3.97~mas.
This value of \thfl\ would suggest that SN~1986J has {\em shrunk}\/
over the last six years, down from \thfl\ = 4.23 mas at $t = 25.6$~yr.
The reason for the apparent decrease in size is that \thfl\ is no
longer a good estimate of the extent of the shell.  As can be seen in
Figure~\ref{fimg}, the extended shell emission is now only just
visible, and the image is strongly dominated by the central component.
Since the central component is at best marginally resolved, an
increase in its brightness will cause \thfl\ to shrink.  We attempted
to compensate for the dominance of the central component by
calculating \thfl\ from an image with an artificially limited
brightness.  This procedure indeed results in larger values of \thfl,
but the exact value is dependent on the surface brightness cutoff
used, and, as the cutoff is lowered, the level of the adjusted
\thfl\ contour becomes comparable to the background rms, and the value
of \thfl\ therefore unreliable.

We therefore can no longer determine the outer radius of SN~1986J due
to the low signal-to-noise ratio of the shell emission.  Our new VLBI
image suggests stronger deceleration since 2008, but a continued
powerlaw expansion with $r \propto t^{0.69}$ is also compatible with
our VLBI image.  There does however, seem to be an evolution of the
radial distribution of the emission, with the emission near the
outside edge having faded relative to that near the center in the last
image.

\section{Discussion}
\label{sdiscuss}

The general picture of radio emission in SNe \citep[see,
  e.g.,][]{Chevalier1982b, ChevalierF2016} is that it arises from the
shocks formed as the cloud of ejecta interacts with the surrounding
circumstellar material (CSM). In particular, a forward shock is driven
into the CSM, a reverse shock is driven back into the expanding
ejecta.  The radio emission is thought to arise from shock-accelerated
electrons and amplified magnetic field between these two shocks.  Both
shocks decelerate with time as the ejecta transfer part of their
kinetic energy to the swept-up CSM\@.  In the case of spherical
symmetry and power-law density distributions, with the CSM density,
$\rhoCSM \propto r^s$ and that of the ejecta, $\rhoeject \propto r^n$,
a self-similar solution exists, and the shock radii are expected to
evolve $\propto r^m$ where $m = (n-3)/(n-s)$ \citep{FranssonLC1996}.
In this picture, one expects radio emission from a spherical shell
region.  In such a region, the outer bound of the radio emission
corresponds to the projected outer edge of the shell which is the
forward shock.  If the volume emissivity is uniform, the brightness
distribution is highest at the projected inner radius of the shell,
and has a minimum in the center.  In the case of SN~1993J, indeed, a
textbook shell structure is seen in the VLBI images
\citep[e.g.,][]{SN93J-3, Marcaide+1995a,Marti-Vidal+2011a}

In SN~1986J, the structure is not as clear, with the shell being
somewhat distorted.  If we equate the outer edge of the radio emission
region with the forward shock\footnote{As we noted in
  \citetalias{SN86J-2}, the angular radius of the forward shock is
  probably not identical to \thfl, our estimate of the outer edge of
  the radio emission, but we expect the two to be close, and moreover,
  since we used a convolving beam that scales with the expansion, for
  the ratio between \thfl\ and the shock radius to remain relatively
  constant as the SN expands.}, we could determine that $m = 0.69 \pm
0.03$ up to $t = 25.6$~yr \citepalias{SN86J-2}.  The outer edge of the
radio emission moved outwards between $t = 0$ and 25.6~yr with an
average speed of 7800~\kms, and a speed at $t = 25.6$~yr of 5400~\kms.
In the present, $t = 31.6$~yr, image, the outer edge of the radio
emission is poorly determined due to the low signal-to-noise ratio.
Although continued expansion with $r \propto t^{0.69}$ is compatible
with our measurements, it requires that the emission near the forward
shock is fading.  In other words, the locus of the brightest radio
emission seems to have stopped expanding and may even be moving
inward.  The radio emission region, even aside from the emergence of
the central component seems to be evolving in a non self-similar
fashion.  It is interesting to note that \citet{DwarkadasG2012} found
that also the observed X-ray emission of SN~1986J was hard to
reconcile with a self-similar evolution since early times.

The central component is stationary to within our uncertainties of
570~\kms, corresponding to $8$\% of the average expansion speed of the
shell.  It is marginally resolved, and has a radius (HWHM) of
$r_\scomp = 6.7_{-3.7}^{+0.7} \times 10^{16}$~cm.  If it had radius =
0 at the time of the explosion in 1983.2, its average expansion speed
since then was $680^{+80}_{-380}$~\kms, or $9^{+1}_{-5}$\% of that of
the shell.

The motions of both the central component and the shell hot-spot, and
the expansion of the more diffuse shell emission, are all therefore
compatible with a homologous expansion with $r \propto t^{0.69}$,
originating from the geometrical center of the shell, which is
coincident within the uncertainties with the position of the central
component.  However, the expansion does {\em not}\/ in fact seem to be
entirely homologous, since the emission near the outer edge seems to
be fading relative to that nearer the center.

What is the nature of the central component?  There are several
hypotheses.  We suggested in \citetalias{SN86J-2} that it could be
emission from the supernova shock running into a dense condensation in
the CSM fortuitously near the center of the shell in projection.  As
mentioned in \S~\ref{sfluxd}, the brightness of the central component
is at least $10\times$ brighter than the shell.  This high and still
increasing brightness, coupled with the central component's
stationarity, and its long lifetime argue against that hypothesis.  It
seems likely therefore that the central component really is in or near
the three-dimensional center of SN~1986J.

In this case the central component may be due to the supernova shock
interacting with a highly structured CSM produced by a binary
companion \citep[Chevalier 2012; Chevalier 2014; see
  also][]{BarkovK2011, PapishSB2015}, \nocite{Chevalier2012,
  Chevalier2014}
where the shock travels much more slowly in the denser parts of the
CSM near the binary orbit plane, thus producing a bright but compact
radio emission region.

The central component could also be emission from a compact remnant of
the supernova.  If that remnant is a neutron star, the emission could
be from a pulsar-wind nebula or perhaps from the neutron star's
accretion disc.  If the remnant is a black hole, which quite possible
given the probably massive progenitor \citep{WeilerPS1990},
then the central component could be emission from the black hole's
accretion disc. In this respect, the coincidence (mentioned in
\S~\ref{sshellspot}) of the approximate alignment of the elongation of
the central component with the NE hot spot and the possible SW
enhancement becomes intriguing. Although inconclusive, it is
suggestive of the action of a jet and counterjet, which has been
suggested as an alternate to the delayed neutrino shock mechanism for
core collapse SNe \citep[e.g.,][]{Soker2010, GilkisSP2016}.  We will
discuss the nature of the central component in more detail in a
forthcoming Paper IV on the evolution of the spectral energy
distribution.

\section{Summary and Conclusions}

\begin{trivlist}

\item{1.} We obtained a new phase-referenced global-VLBI image of
  SN~1986J at 5~GHz, showing the continued evolution of this supernova
  in the radio.  An animation is available in the electronic edition.

\item{2.} The 5-GHz VLBI image is now dominated by a marginally
  resolved central component.  The ejecta shell is only barely
  visible.  The peak brightness of the central component is $\sim$10
  times higher than that of the shell.
  
\item{3.} The flux density of the central component is still
  increasing, both in absolute terms and as a fraction of the total.
  Since its first detection in 2003 ($t=20.3$~yr), its 5-GHz flux
  density has increased by a factor of $\sim$4.  Its current
  luminosity ($\nu\,L_\nu$ at 5~GHz) is $20 \sim 30$ times that of the
  Crab Nebula ($3.5 \sim 4.5 \times 10^{35}$ erg~s$^{-1}$).
  % Use 590 Jy/5GHz for Crab; 511 uJy = 21.2 Crab, 750 uJy = 31.8 Crab

\item{4.} The $3\sigma$ upper limit on the linearly polarized fraction
  of the image peak, which is the central component, was 3.3\%

\item{5.} The emission from the shell is decreasing, and the
  brightness of the outer edge seems to have faded more than that
  nearer the center.  The angular outer radius of the radio emission,
  identified with the location of the forward shock, is no longer well
  determined.  Continued expansion with $r \propto t^{0.69}$ seen
  earlier is consistent with our new VLBI image, but it is also
  possible that the deceleration has increased.

\item{6.} In earlier observations, there was a prominent hot-spot to
  the NE in the shell. The shell hot-spot, which has almost faded from
  view in our present, $t=31.6$~yr, image, was moving outward with a
  projected speed of $2810 \pm 750$~\kms\ at p.a.\ $57\arcdeg \pm
  15\arcdeg$ between $t = 15.9$ and 25.6 yr.  The shell hot spot's
  motion is consistent with it taking part in a homologous expansion
  together with the shell, in other words, having an origin at the
  explosion center in 1983.2 and moving radially outward with $r
  \propto t^{0.69}$.

\item{7.} The central component seems to be marginally resolved in our
  observations.  We found its FWHM angular diameter at $t = 31.6$~yr
  to be $900_{-500}^{+100}$~\muas, corresponding to $r_\scomp =
  6.7_{-3.7}^{+0.7} \times 10^{16}$~cm.  If it has expanded since the
  explosion in 1983.2, then the average projected speed of expansion
  was $680^{+80}_{-380}$~\kms, or 9\% the speed of the outer edge of
  the shell.

\item{8.} The central component has a proper motion corresponding to a
  projected speed of $380 \pm 570$~\kms\ between $t = 20.3$~yr and
  31.6~yr (2003 and 2014), consistent with being stationary.  Its
  position is consistent within the uncertainties of the explosion
  position that we estimated earlier.

\item{9.} The latest observations argue in favor of the central
  component being located at or near the three-dimensional center of
  SN~1986J, rather then being associated with the expanding shell and
  being central only in projection.

\end{trivlist}

\section*{Acknowledgments }

We thank N. Soker for comments on the manuscript.  The European VLBI
Network is a joint facility of European and Chinese radio astronomy
institutes funded by their national research councils.  This research
was supported by both the National Sciences and Engineering Research
Council of Canada and the National Research Foundation of South
Africa.  We have made use of the NASA/IPAC Extragalactic Database
(NED) which is operated by the Jet Propulsion Laboratory, California
Institute of Technology, under contract with the National Aeronautics
and Space Administration (NASA), as well as NASA's Astrophysics Data
System Abstract Service.
\bibliographystyle{apj} 
\bibliography{mybib1}

\clearpage

\end{document}